\documentclass[sn-aps,iicol,pdflatex]{sn-jnl}

\usepackage{graphicx}%
\usepackage{multirow}%
\usepackage{amsmath,amssymb,amsfonts}%
\usepackage{amsthm}%
\usepackage{mathrsfs}%
\usepackage[title]{appendix}%
\usepackage{xcolor}%
\usepackage{textcomp}%
\usepackage{manyfoot}%
\usepackage{booktabs}%
\usepackage{algorithm}%
\usepackage{algorithmicx}%
\usepackage{algpseudocode}%
\usepackage{listings}%
\usepackage{graphicx}%
\usepackage{multirow}%
\usepackage{amsmath,amssymb,amsfonts}%
\usepackage{amsthm}%
\usepackage{mathrsfs}%
\usepackage[title]{appendix}%
\usepackage{xcolor}%
\usepackage{textcomp}%
\usepackage{manyfoot}%
\usepackage{booktabs}%
\usepackage{algorithm}%
\usepackage{algorithmicx}%
\usepackage{algpseudocode}%
\usepackage{listings}%
\usepackage{bm,physics}
\usepackage{amsmath,amssymb}
\usepackage{ulem, color}
\usepackage{amsmath}
\usepackage{natbib}
\bmdefine{\ba}{a}
\bmdefine{\bb}{b}
\bmdefine{\bx}{x}
\bmdefine{\by}{y}
\bmdefine{\bz}{z}
\bmdefine{\bn}{n}
\bmdefine{\bp}{p}
\newcommand{\BM}{\begin{pmatrix}}
\newcommand{\EM}{\end{pmatrix}}
\renewcommand{\d}{\dagger}

\newcommand{\Lc}{\mathcal{L}}
\newcommand{\Mc}{\mathcal{M}}
\newcommand{\hphi}{\hat\varphi}
\newcommand{\hpsi}{\hat\psi}

\newcommand{\ex}{\mathrm{ex}}

\newcommand{\REVR}[1]{\textcolor{black}{{\bf #1}}} 
\newcommand{\REV}[1]{\textcolor{black}{{ #1}}}  
\newcommand{\REVB}[1]{\textcolor{black}{{ #1}}}

\begin{document}

\title {Six-$\alpha$ cluster Bose-Einstein  condensation and supersolid  $^{12}$C($0_2^+)$+$^{12}$C($0_2^+)$ molecular  structure in $^{24}$Mg  
}
\author*[1]{\fnm{S.} \sur{Ohkubo}}\email{ohkubo@rcnp.osaka-u.ac.jp}

\author[2]{\fnm{J.} \sur{Takahashi}}

\author[3]{\fnm{Y.} \sur{Yamanaka}}

\affil*[1]{\orgdiv{Research Center for Nuclear Physics}, \orgname{Osaka University}, \orgaddress{\street{} \city{Ibaraki}, \postcode{567-0047}, \state{} \country{ Japan}}}

\affil[2]{\orgdiv{Faculty of Economics}, \orgname{Asia University}, \orgaddress{\street{} \city{ Tokyo}, \postcode{ 180-0022}, \state{} \country{Japan}}}

\affil[3]{\orgdiv{Institute of Condensed-Matter Science}, \orgname{Waseda University}, \orgaddress{\street{} \city{ Tokyo}, \postcode{ 169-8555}, \state{} \country{Japan}}}

\abstract{We show for the first time that the low-spin ($J \le 4^+$) six-$\alpha$ condensate candidate states in $^{24}$Mg, recently reported by Fujikawa et al. [Phys. Lett. B 848, 138384 (2024)], are well described by the superfluid $\alpha$-cluster model (SCM). This is achieved by a rigorous treatment of the Nambu-Goldstone (NG) zero mode as the order parameter of condensation in the finite six-$\alpha$ system. 
We find that a roton rotational band with a large moment of inertia is built on the first excited NG $0^+$ state, analogous to the roton bands observed in three-, four-, and five-$\alpha$ condensates in $^{12}$C, $^{16}$O, and $^{20}$Ne, respectively. Remarkably, our calculated roton band reproduces the well-known molecular resonance with a $^{12}$C($0_2^+$)+$^{12}$C($0_2^+$) structure ($16^+$) observed at $E_{\rm c.m.} = 32.5$ MeV in inelastic $^{12}$C+$^{12}$C scattering. This result provides a unified description of both the low-spin six-$\alpha$ condensate states and the high-spin $^{12}$C($0_2^+$)+$^{12}$C($0_2^+$) molecular resonance. Analysis of the wave functions reveals a large overlap between the SCM states and a geometrical $^{12}$C($0_2^+$)+$^{12}$C($0_2^+$) configuration. This dual nature —the coexistence of superfluidity and crystallinity— identifies these states as a signature of a supersolid.
}

\keywords{six-$\alpha$  condensation in $^{24}$Mg,  superfluid cluster model,\REV{ roton band}, $^{12}$C($0_2^+)$+$^{12}$C($0_2^+)$ \REV{cluster structure},  supersolidity}

\maketitle

\section{Introduction}\label{sec:1}
Very recently, experimental six-$\alpha$ cluster candidate states in $^{24}$Mg were reported by Fujikawa et al. \cite{Fujikawa2024}.
Following the observation of Bose-Einstein condensation (BEC) in trapped cold atoms \cite{Anderson1995,Cornel2002}, $\alpha$-cluster condensation in light nuclei such as $^{12}$C \cite{Tohsaki2001,Matsumura2004,Funaki2015,Funaki2016,Nakamura2016,%
Katsuragi2018,Itoh2011A,Itoh2011B,Freer2009,Zimmerman2011,Zimmerman2013,%
Itoh2013,Freer2011}, $^{16}$O \cite{Tohsaki2001,Funaki2008C,Ohkubo2010,Funaki2018,Takahashi2020%
,Freer1995,Freer2004,Freer2005,Itoh2014,Curtis2016,Barbui2018}, and $^{20}$Ne \cite{Swartz2015,Adachi2020,Zhou2023,Ohkubo2025} has been extensively investigated both theoretically and experimentally in recent decades.

\par
Although the developments of microscopic superfluidity in cold atoms \cite{Otani2025}—where superfluidity due to BEC has been known to exist even in small systems with ten or fewer interacting trapped particles \cite{Kuyanov2008,Khairallah2007,Li2010,Paesani2005,McKellar2006}—give support to the idea of BEC of a few to several $\alpha$-cluster condensates in light nuclei, the observation of typical superfluid phenomena \cite{Brink2005} would definitively confirm the existence of BEC $\alpha$-cluster structures. These phenomena include roton excitations \cite{Landau1941,Landau1947,Feynman1953A,Feynman1953B,Feynman1954,Feynman1957}, quantum vortices \cite{Cornel2002}, or the Josephson effect \cite{Josephson1966,Broglia2022}.

\par
The six-$\alpha$ condensate candidate states were observed at center-of-mass energies ($E_{\rm c.m.}$) of 19.3 MeV, 21.4 MeV ($2^+$), and 22.5 MeV ($4^+$) from the $^{12}$C+$^{12}$C threshold in the $^{12}$C+$^{12}$C → $^{12}$C($0_2^+$) (Hoyle state)+X reactions \cite{Fujikawa2024}. These states are located slightly above the $^{12}$C($0_2^+$)+$^{12}$C($0_2^+$) threshold energy in $^{24}$Mg.
Six-$\alpha$  condensation in $^{24}$Mg is particularly interesting because both $\alpha$ -cluster and two Hoyle-like molecular $^{12}$C+$^{12}$C structures appear in the relevant excitation energy region. The latter has been the most thoroughly studied, pioneering the field of nuclear molecular structure \cite{Bromley1960,Bromley1984,Betts1997}.

\par
On the other hand, regarding the six-$\alpha$ cluster structure versus the molecular $^{12}$C+$^{12}$C structure, there was an intensive theoretical and experimental investigation into a six-$\alpha$ cluster state in the highly excited energy region of $^{24}$Mg more than thirty years ago. This search was for a six-$\alpha$ \REVB{linear} chain structure \cite{Horiuchi1972,Marsh1986,Rae1992}, an analog to the three-$\alpha$  \REVB{linear} chain structure in $^{12}$C \cite{Morinaga1956} and a four-$\alpha$ linear chain structure in $^{16}$O \cite{Chevallier1967}.

\par
The observation of a six-$\alpha$ cluster state at $E_{\rm c.m.}$= 32.5 MeV in inelastic $^{12}$C+$^{12}$C scattering \cite{Wuosmaa1992,Chappell1995} sparked controversial discussions. The state was assigned a spin-parity of $J^\pi=16^+$ or $14^+$ in Ref. \cite{Wuosmaa1992} and later independently confirmed as $16^+$ by Szilner et al. \cite{Szilner1997}. The debate was over a six-$\alpha$ linear chain structure  \cite{Rae1992,Rae1995} versus a  $^{12}$C($0_2^+$)+$^{12}$C($0_2^+$) structure. Quantitative analysis \cite{Hirabayashi1995, Chappell1995} of the experimental data showed that the observed state was a molecular state with the $^{12}$C($0_2^+$)+$^{12}$C($0_2^+$) structure rather than a six-$\alpha$ linear chain.

\par
In the early 1990s, before the discovery of Bose-Einstein condensation in dilute cold atoms \cite{Anderson1995,Cornel2002}, no one considered the observed 32.5 MeV state with the $^{12}$C($0_2^+$)+$^{12}$C($0_2^+$)  structure in light of BEC of six $\alpha$ clusters. This was because the Hoyle state was still believed to have a three-$\alpha$ chain structure, even though its gas-like, weakly coupled three-$\alpha$ nature had already been revealed by Uegaki et al. \cite{Uegaki1977,Uegaki1978,Uegaki1979}.

\par 
 A recent finding of a roton band, \REV{which is a rotational band formed by excitation  with a finite angular momentum  of the superfluid vacuum} 
 \REV{(roton  \cite{Landau1941,Landau1947,Feynman1953A,Feynman1953B,Feynman1954,%
 Feynman1957}) on top of
 the Nambu-Goldstone (NG) mode (phonon mode)},
  with supersolidity in the $\alpha$ condensates of $^{12}$C, $^{16}$O, and $^{20}$Ne \cite{Ohkubo2025} suggests that this supersolidity may also persist in the six-$\alpha$ condensate related to the $^{12}$C+$^{12}$C molecular structure in $^{24}$Mg.
Therefore, it is both important and challenging to study the three six-$\alpha$ condensate candidate states with low spins observed just above the six-$\alpha$ threshold in Ref. \cite{Fujikawa2024} and the \REVB{well-known} 32.5 MeV (16$^+$) molecular state with the $^{12}$C($0_2^+$)+$^{12}$C($0_2^+$) structure at high excitation energy in Refs. \cite{Wuosmaa1992,Chappell1995,Szilner1997} in a unified way.
It is also particularly interesting to reveal whether the observed $\alpha$ condensate candidate states are related to the $^{12}$C($0_2^+$)+$^{12}$C($0_2^+$) structure and whether they are persistently related to the roton band with supersolidity of a six-$\alpha$ condensate, similar to the findings in $^{12}$C, $^{16}$O, and $^{20}$Ne \cite{Ohkubo2025}.

\par
The purpose of this paper is to show that the observed six-$\alpha$ cluster states can be reproduced by the superfluid cluster model (SCM), and that a BEC rotational roton band with six $\alpha$ clusters is built just above the six-$\alpha$ cluster vacuum. The mechanism of roton excitation is found to be logically identical to that for $^{12}$C, $^{16}$O, and $^{20}$Ne.
The rotational band is shown to exhibit the dual property of superfluidity and crystallinity—a property of a supersolid—and is also well-described as having a geometrical $^{12}$C($0_2^+$)+$^{12}$C($0_2^+$) cluster structure.
Surprisingly, we find that the observed six-$\alpha$ cluster state with the molecular $^{12}$C($0_2^+$)+$^{12}$C($0_2^+$) structure at $E_{\rm c.m.}$=32.5 MeV can be described by the SCM. We also find that the six-$\alpha$ condensate states with low spins slightly above the vacuum and the $^{12}$C($0_2^+$)+$^{12}$C($0_2^+$)  molecular resonances are understood in a unified way.

\par 
The paper is organized as follows. Section 2 presents the formulation of our field-theoretical superfluid cluster model. Section 3 studies six-$\alpha$ condensation in $^{24}\text{Mg}$ using the SCM and demonstrates the existence of a roton rotational band. Section 4 is devoted to the six-$\alpha$ condensation and the $^{12}$C($0_2^+$)+$^{12}$C($0_2^+$)  molecular resonance in $^{24}\text{Mg}$. Supersolidity with a \REVB{ $^{12}$C($0_2^+$)+$^{12}$C($0_2^+$)}  dinuclear Bose-Einstein condensate \REVB{of the roton band states}  is demonstrated in Sect. 5. 
Finally, Section \REV{6} provides a summary.

\section{The field theoretical six-$\alpha$ superfluid cluster model for  $^{24}$Mg}\label{sec:2}
\par
The traditional $\alpha$-cluster models have no superfluidity order parameter, which makes it impossible to conclude whether the BEC candidate states in $^{12}$C, $^{16}$O, and $^{20}$Ne are truly BEC. Also, traditional cluster models have not yet explained the BEC nature of the observed rotational band states built just above the BEC vacuum state near the $\alpha$-breakup threshold.

\par
In order to understand the collective motions of $\alpha$ clusters built on the BEC vacuum, it is essential to use a theory with an order parameter that rigorously treats the Nambu-Goldstone (NG) mode caused by the spontaneous symmetry breaking (SSB) of the global phase. The present superfluid cluster model (SCM) rigorously treats the NG mode and has been successfully applied to reveal the BEC nature and supersolidity of three-, four-, and five-$\alpha$ condensations in $^{12}$C \cite{Nakamura2016,Katsuragi2018,Ohkubo2020B}, $^{16}$O \cite{Takahashi2020}, and $^{20}$Ne \cite{Ohkubo2025}, respectively. It was also applied to supersolidity and soft modes in heavier systems such as $^{40}$Ca \cite{Ohkubo2022} and $^{52}$Fe \cite{Katsuragi2018}.

\par
We briefly recapitulate the formulation  of the SCM \cite{Nakamura2014,Nakamura2016}.
The model Hamiltonian for a bosonic field $\hpsi(\REV{\bf r})$
representing
the $\alpha$ cluster is given as follows:
\begin{align}
&\hat{H}=\int d\REV{\bm r} \hpsi^\d(\REV{\bm r}) \left(-\frac{\nabla^2}{2m}+
V_\ex(\REV{\bm r})- \mu \right) \hpsi(\REV{\bm r}) \notag\\
&\,\,+\frac12 \int d\REV{\bm r}d\REV{\bm r'} \hpsi^\d(\REV{\bm r})
\hpsi^\d(\REV{\bm r}') U(|\REV{\bm r}-\REV{\bm r}'|) \hpsi(\REV{\bm r}') \hpsi(\REV{\bf r}) \,.
\label{Hamiltonian}
\end{align}
Here, where $m$ and $\mu$ denote the mass of the $\alpha$ particle and the chemical
potential, respectively. The potential $V_\ex$ is introduced phenomenologically 
to trap the $\alpha$ clusters inside the nucleus, and is taken to have a harmonic
form,
$
V_\ex(r)= m \Omega^2 r^2/2\,,
$
and the $\alpha$--$\alpha$ interaction
is given by the Ali--Bodmer potential \cite{Ali1966},
$
U (|\bm r -\bm r'|) = V_r e^{-\mu_r^2 |\bm r -\bm r'|^2}
- V_a e^{-\mu_a^2 |\bm r -\bm r'|^2}\,.
$
We set $\hbar=c=1$ throughout this paper.

\par
When BEC of $\alpha$ clusters occurs, i.e.
the global phase symmetry of $\hpsi$ is spontaneously broken, we decompose $\hpsi$
as $\hpsi(\REV{\bm r})=\xi(\REV{\bm r})+\hphi(\REV{\bm r})$, where the c-number $\xi(r)=\bra{0} \hat\psi(\REV{\bm r})
\ket{0}$ is an order parameter and is assumed to be real and isotropic.
To obtain the excitation spectrum, we need to solve
three coupled equations, which are the Gross--Pitaevskii (GP) equation, Bogoliubov-de Gennes (BdG) equations, and Nambu-Goldstone zero-mode equation \cite{Nakamura2016, Katsuragi2018}.
The GP equation determines the order parameter $\xi$ by
\begin{equation}\label{eq:GP}
\left\{ -\frac{\nabla^2}{2m}+V_\ex(r) -\mu + V_H(r)
\right\} \xi(r) = 0 \,,
\end{equation}
where
$ V_H(r) = \int d\REV{\bm r'} U(|\REV{\bm r}-\REV{\bm r'}|)\xi^2(r')$. 
$\xi$ is normalized
with the condensed particle number $N_0$ as
\begin{align}
\int d\REV{\bm r} |\xi(r)|^2 = N_0\,.
 \end{align}
The superfluidity density is given by  $\rho(r)$=$|\xi(r)|^2/N_0$\,.
The BdG equations describe the collective motions on the condensate
by
\begin{align}
\int d\REV{\bm r'}
\left(\begin{array}{cc}
\Lc & \Mc \\
-\Mc^* & -\Lc^*
\end{array}\right)
\left(\begin{array}{c}
u_{\bn} \\
v_{\bn}
\end{array}\right)
= \omega_\bn
\left(\begin{array}{c}
u_{\bn} \\
v_{\bn}
\end{array} \right),
\end{align}
where
$\Mc(\REV{\bm {r}}, \REV{\bm r'}) 
=  U(|\REV{\bm r}-\REV{\bm r'}|) \xi(r) \xi(r'),\, 
\Lc{\textcolor{black}(}\REV{\bm r}, \REV{\bm r'})
= \delta{\textcolor{black}(}\REV{\bm r}-\REV{\bm r'})
\left\{ -\frac{\nabla^2}{2m}+V_\ex(r) -\mu + V_H(r)\right\}
+ \Mc{\textcolor{black}(}\REV{\bm r}, \REV{\bm r'})\,. 
$
The index $\bm{n}=(n,\, \ell,\, m)$ stands for the main, azimuthal and
magnetic quantum numbers. The eigenvalue $\omega_\bn$ is the excitation energy
of the Bogoliubov mode. For isotropic $\xi$, the BdG eigenfunctions can be taken to have separable forms,
\begin{align}
\label{eq:UV}
u_{\bm{n}}(\bm{r}) = \mathcal{U}_{n\ell}(r) Y_{\ell m}(\theta, \phi), \, \nonumber \\
v_{\bm{n}}(\bm{r}) = \mathcal{V}_{n\ell}(r) Y_{\ell m}(\theta, \phi).
\end{align}
We necessarily have an eigenfunction belonging
to zero eigenvalue, explicitly $(\xi(r), -\xi(r))^t$, and its adjoint function
$(\eta(r),\eta(r))^t$ is obtained as
$\eta(r)= \frac{\partial}{\partial N_0}\xi(r)\,.$
The field operator is expanded as
$\hphi(\REV{\bm r})=-i{\hat Q}\xi(r)+{\hat P}\eta(r) 
+\sum_{\bn} \left\{{\hat a}u_\bn(\REV{\bm r})
+{\hat a}^\dagger v^\ast_\bn(\REV{\bm r}) \right\}\,,$
with the commutation relations $[{\hat Q}\,,\,{\hat P}]=i$ and
$[{\hat a}_{\bn}\,,\,{\hat a}^\dagger_{\bn'}]=
\delta_{\bn \bn'}$\,. The operator ${\hat a}_\bn$ is an annihilation operator
of the Bogoliubov mode, and the pair of canonical operators ${\hat Q}$ and ${\hat P}$
originate from the SSB of the global phase and are called the NG zero-mode operators.

\par
The treatment of the NG zero-mode operators is a chief feature of our approach.
The naive choice of the unperturbed bilinear Hamiltonian with respect
to ${\hat Q}$ and ${\hat P}$ fails due to their large quantum fluctuations.
Instead, we gather all the terms consisting only of ${\hat Q}$ and ${\hat P}$
in the total Hamiltonian to construct the unperturbed nonlinear Hamiltonian
of ${\hat Q}$ and ${\hat P}$, denoted by $H_u^{QP}$\,.
The coefficients in $\hat{H}_u^{QP}$ are $I=\frac{\partial \mu}{\partial N_0}$
and the integrations involving $\xi\,,\,\eta\,$ and $U$, whose explicit forms are
seen in the Ref.~\cite{Katsuragi2018}. We set up the eigenequation
for $\hat{H}_u^{QP}$, called
the NG zero--mode equation,
\begin{align} \label{eq:HuQPeigen}
\hat H_u^{QP} \ket{\Psi_\nu} = E_\nu \ket{\Psi_\nu}\qquad
(\nu=0,1,\cdots)\,.
\end{align}
This equation is similar to a one-dimensional Schr\"odinger equation for a bound problem. 

\par
The total unperturbed Hamiltonian ${\hat H}_u$ is ${\hat H}_u=\hat H_u^{QP}
+\sum_{\bn} \omega_\bn {\hat a}_\bn^\dagger{\hat a}_\bn$.
The ground state (vacuum) energy is set to zero, $E_0=0$.
The states that we consider are $\ket{\Psi_\nu}\ket{0}_{\rm ex}$ with
energy $E_\nu$, called the NG zero-mode state, and
$\ket{\Psi_0}{\hat a}^\dagger_\bn
\ket{0}_{\rm ex}$ with energy $\omega_\bn$, called the BdG state, 
where ${\hat a}_\bn \ket{0}_{\rm ex}=0$.

\section{Six-$\alpha$  condensation in $^{24}$Mg and roton rotational band}\label{sec:3}

\par
The two parameters    $\Omega$ and $V_r$ control the size and stability of the trapped condensate, respectively. They  are determined  as in the case for $^{12}$C, $^{16}$O, and  $^{20}$Ne in  Refs.~\cite{Nakamura2016,Katsuragi2018,Takahashi2020,Ohkubo2025}. 
The parameter $\Omega$ \REV{is }  determined to reproduce the size of the condensate vacuum, around $R_0$=6.4 fm, which is  scaled from the sizes  of the condensate, $\sim$4.2 fm for  $^{12}$C, $\sim$5.6 fm  for $^{16}$O,  and   $\sim$6.0 fm for $^{20}$Ne.
     $V_r$ is determined to \REV{reproduce the  excitation energy of 22.5 MeV of the  $2^+$ state  observed as a six-$\alpha$  condensate candidate  in Ref. \cite{Fujikawa2024}. The determined values are  $\Omega$=1.3305 MeV/$\hbar$ and $V_r$= 558 MeV.
   }
 The chemical potential is fixed by the input $N_0$.
 \REV{The condensation rate ($R_{\rm con}$) is defined as $R_{\rm con}=N_0/N$, where $N_0$ is the number of $\alpha$ clusters occupying the lowest-energy $0s$ state (i.e., not excited) and $N=6$ is the total number of $\alpha$ clusters in the system.
}

\par
\begin{figure}[t!]
\begin{center}
\includegraphics[width=7.8cm]{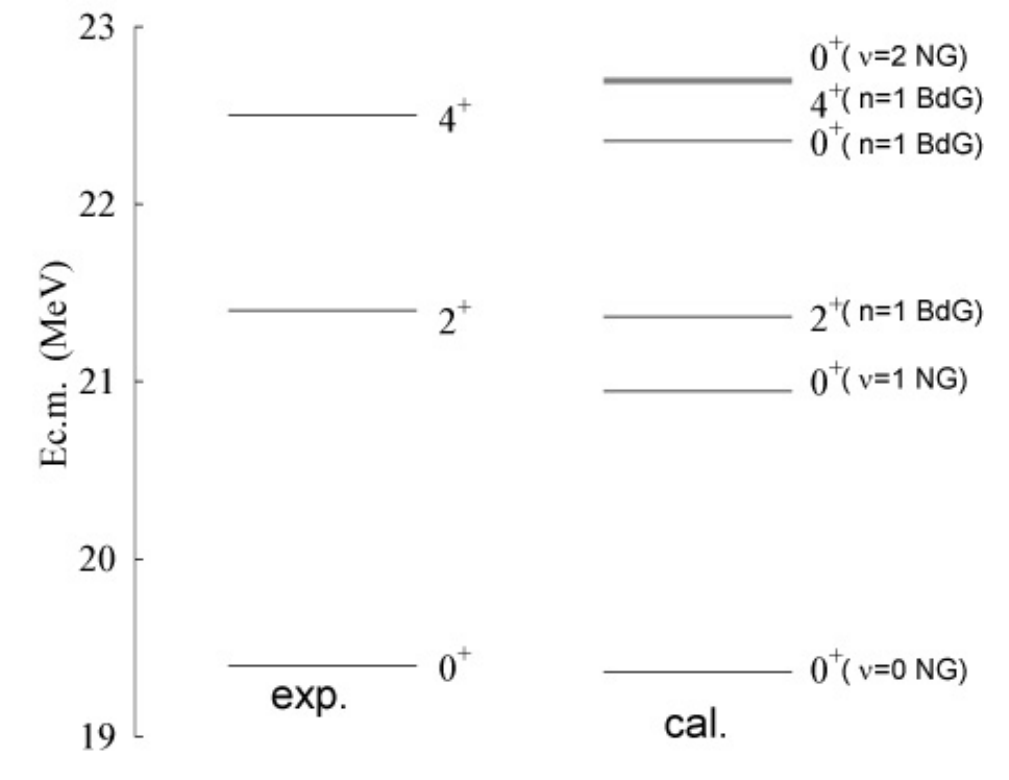}
\caption{  The experimental energy levels of the six-$\alpha$ condensate candidate states in $^{24}$Mg \cite{Fujikawa2024} are compared with the SCM calculations with a condensation rate of 70\%. The vertical axis shows the energy from the $^{12}$C+$^{12}$C threshold in $^{24}$Mg
   }
\label{fig1}
\end{center}
\end{figure}

In Fig.~\ref{fig1}, the energy levels calculated with the \REV{above determined
$\Omega$ and  $V_r$}  are displayed. In the calculations, the observed state at $E_{\rm c.m.}$=19.4 MeV ($E_{\rm x}$=33.3 MeV in $^{24}$Mg)  is assumed to be a $0^+$ vacuum state.
 Our calculations predict a $0^+$ state as the first excited state of a six-$\alpha$ condensate. Since the first excited state is a member of the NG mode caused by the SSB of the vacuum's global phase, this mechanism  is robust as seen in the 
three-$\alpha$,  four-$\alpha$, and five-$\alpha$ condensation in 
$^{12}$C \cite{Itoh2011A,Itoh2011B,Nakamura2016,Katsuragi2018}, $^{16}$O  \cite{Itoh2014,Takahashi2020}, and $^{20}$Ne \cite{Adachi2020,Ohkubo2025}.

\par
\begin{figure}[t!]
\begin{center}
\includegraphics[width=7.8cm]{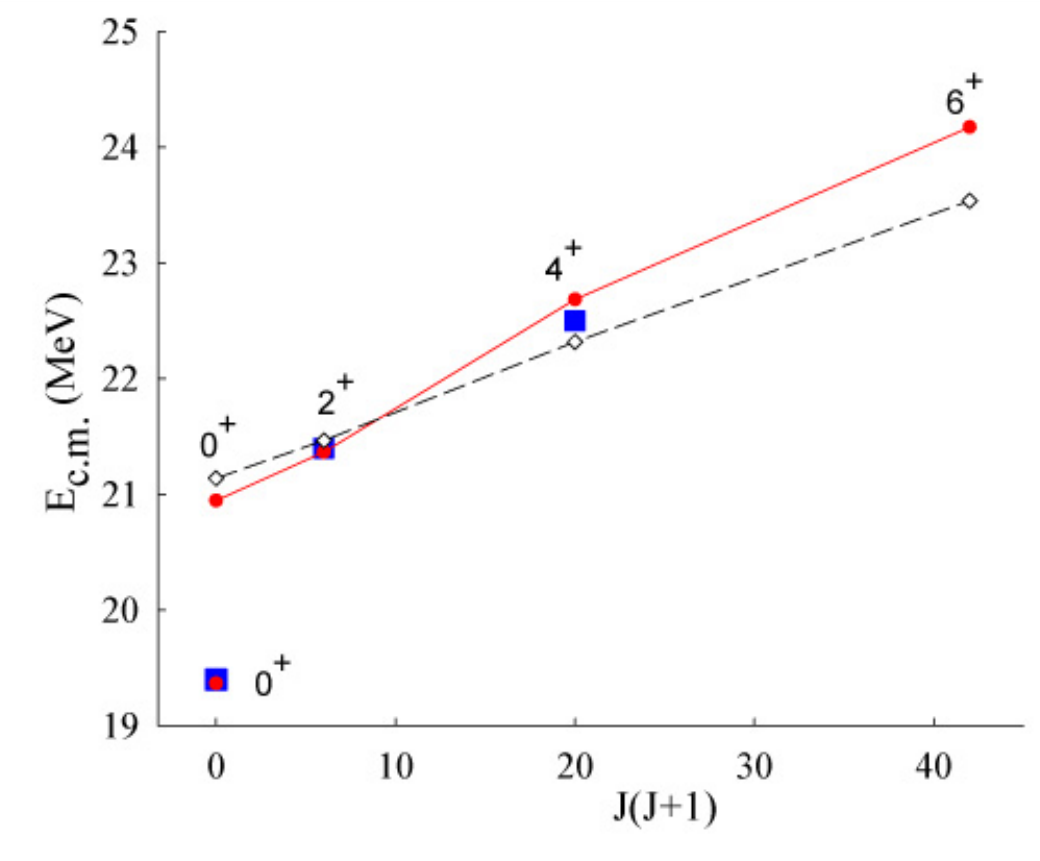}
\caption{ 
The rotational roton band $K_{\rm rot}$   (closed circles), built on the NG mode $0^+$  of the six-$\alpha$  condensate and calculated in the SCM with a condensation rate of 70\%, is compared with the experimental six-$\alpha$  candidate states (squares) in $^{24}$Mg \cite{Fujikawa2024}. This band is excited from the vacuum $0^+$  state near 19 MeV. For comparison, the band of molecular resonances with a $^{12}$C($0_2^+$)+$^{12}$C($0_2^+$)  structure (open diamonds), calculated using the coupled-channel method \cite{Hirabayashi1995} based on the crystallinity picture, is also presented. The lines serve as a guide for the eye for the rotational bands. The vertical axis represents the energy from the  $^{12}$C+$^{12}$C threshold in $^{24}$Mg
}
\label{fig2}
\end{center}
\end{figure}

In Fig.~\ref{fig2}, the observed $2^+$ and  $4^+$  states are plotted against  $J(J+1)$ to clearly understand their nature, whether they are  vibrational states,  rotational states, or in-between. The SCM calculations show a rotational band with a rotational constant $k\approx$78 keV, where $k$=$\frac{\hbar^2}{2\mathcal{J}}$    and $\mathcal{J}$ is the moment of inertia.
In Fig.~\ref{fig2}, cluster model calculations in the coupled-channel method from Ref.~\cite{Hirabayashi1995}, which include the excited states  $2_1^+$, $3^-$, $0_2^+$, and $2_2^+$   of $^{12}$C, are presented for comparison. This model also gives a similar rotational band, which has a dominant $^{12}$C($0_2^+$)+$^{12}$C($0_2^+$) cluster structure and places the band head $0^+$  state at almost the same energy as predicted by the SCM.
Both bands calculated with the SCM and the $^{12}$C+$^{12}$C cluster model (crystallinity picture) agree well with the experimental states. Both models predict a band head $0^+$  state at around  $E_{\rm c.m.}$=21 MeV. Therefore, it is highly desirable to search for this band head $0^+$  state in experiments. The characteristic feature of the band from both the condensation and crystallinity pictures is in line with the similar bands in $^{12}$C, $^{16}$O, and $^{20}$Ne, which are built on the first excited NG $0^+$  state above the vacuum, as revealed in \cite{Ohkubo2025}.

\par 
It is not self-evident that the states created by BdG excitation of the spherical BEC vacuum form a rotational band, because in the SCM, no geometrical configuration of the six $\alpha$ clusters is assumed  {\it a priori}. The obtained results show that the BEC condensates of the band have a deformed structure of six $\alpha$  clusters. Also, because both the SCM and the geometrical cluster model with the $^{12}$C($0_2^+$)+$^{12}$C($0_2^+$)  configuration reproduce the experimental data, 
it is suggested \REV{that this band, which is referred to as a roton band in Ref. \cite{Ohkubo2025} due to its creation by excitation with finite angular momentum, exhibits the seemingly incompatible properties of condensation and a geometrical cluster structure. This is confirmed by examining a much higher spin state, $J^\pi=16^+$, in the next Section \ref{sec:4}.
}

 \section{Six-$\alpha$  condensation and $^{12}$C($0_2^+$)+$^{12}$C($0_2^+$)
  molecular  resonance in $^{24}$Mg}\label{sec:4}

\par
\begin{figure}[t!]
\begin{center}
\includegraphics[width=8.0cm]{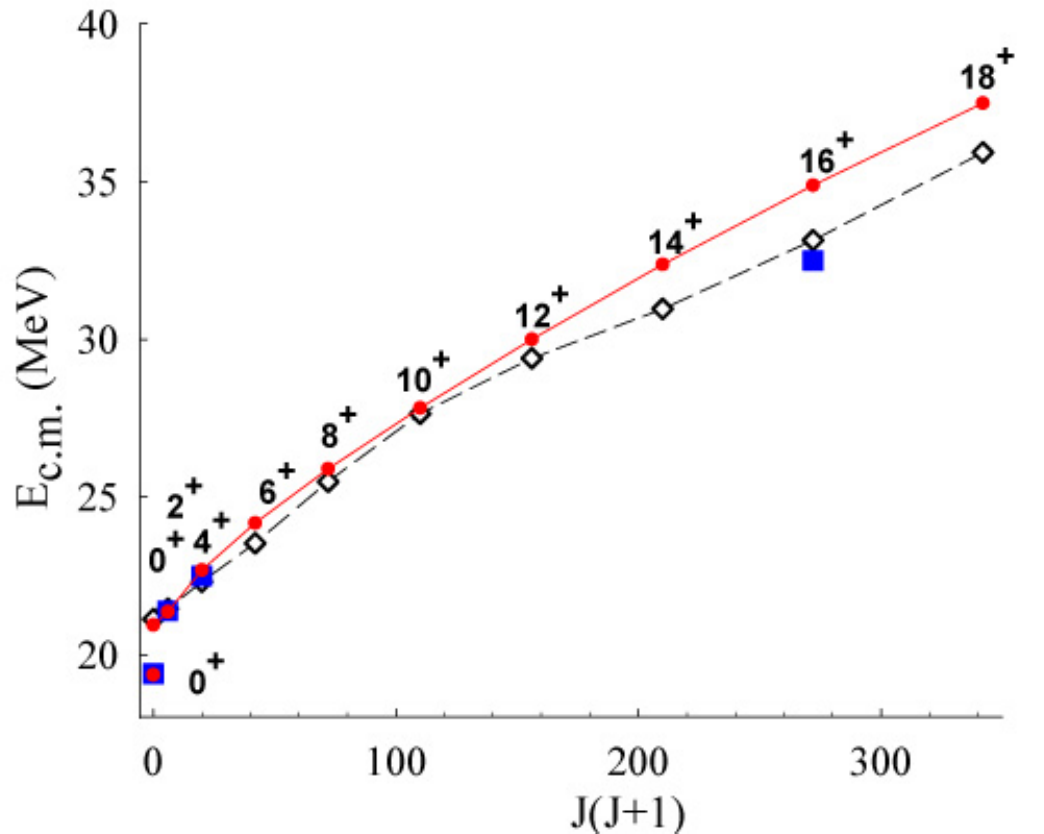}
\caption{
The SCM calculations (closed circles) with a 70\% condensation rate are compared with the experimental data (squares) of the molecular resonance with a high-spin $^{12}$C($0_2^+$)+$^{12}$C($0_2^+$)  cluster structure from Refs.~\cite{Wuosmaa1992,Szilner1997}, as well as the low-spin six-$\alpha$ condensate candidates from Ref.~\cite{Fujikawa2024}  in $^{24}$Mg. For comparison, the theoretical molecular resonances with the  $^{12}$C($0_2^+$)+$^{12}$C($0_2^+$)  cluster structure (diamonds), calculated using the coupled-channel method  \cite{Hirabayashi1995}, are also presented. The lines are a guide for the eye for the rotational roton band $K_{\rm rot}$. The vertical axis represents the energy from the $^{12}$C+$^{12}$C threshold in $^{24}$Mg
}
\label{fig3}
\end{center}
\end{figure}

In Fig.~\ref{fig3}, the band states calculated in the SCM are displayed up to $18^+$. The reason we consider these very high-spin states in the context of six-$\alpha$ condensation is that a state with a $^{12}$C($0_2^+$)+$^{12}$C($0_2^+$) structure was observed by Wuosmaa {et al.} \cite{Wuosmaa1992} in the pioneering measurement  of   $^{12}$C+$^{12}$C  molecular resonance with a six-$\alpha$ cluster structure. This state was found at $E_{\rm c.m.}$=32.5 MeV with a width $\Gamma_{\rm c.m.}$=4.7 MeV in a coincidence experiment. There were discussions about whether this state has a six-$\alpha$ linear chain structure with $k$=22 keV \cite{Rae1992,Marsh1986} or a molecular $^{12}$C($0_2^+$)+$^{12}$C($0_2^+$)  structure with $k\approx$80 keV \cite{Hirabayashi1995}. A quantitative analysis using the coupled-channel method in Ref.~\cite{Hirabayashi1995} reproduced the angular distributions in elastic $^{12}$C+$^{12}$C scattering, as well as inelastic scattering of $^{12}$C($0_2^+$)+$^{12}$C(g.s.)  and $^{12}$C($0_2^+$)+$^{12}$C($0_2^+$), and the excitation functions. This analysis showed that the observed resonance state has a dominant $^{12}$C($0_2^+$)+$^{12}$C($0_2^+$) structure rather than a six-$\alpha$ linear chain structure. The calculation described the observed molecular resonance as a $16^+$ state (see Fig.~2 of Ref.~\cite{Hirabayashi1995}).
\REV{It is also worth noting that this calculation \cite{Hirabayashi1995}, based on the geometrical model, predicts the $^{12}$C(g.s.)+$^{12}$C(g.s.) molecular cluster band to be located in agreement with the experimental observation \cite{Ordonez1986}. The band head $0^+$ is found at around $E_{\rm c.m.}$=7.1 MeV, slightly above the $^{12}$C(g.s.)+$^{12}$C(g.s.) threshold energy.
}

 \par
 It is surprising that the present SCM calculations not only reproduce the low-spin BEC candidate states of Fujikawa et al. \cite{Fujikawa2024} but also that the high-spin $16^+$  state corresponds to the observed molecular state with a  $^{12}$C($0_2^+$)+$^{12}$C($0_2^+$)  cluster structure.
 \REV{It is to be noted that our chosen interaction is not initially optimized to reproduce the high spin states. To reproduce the experimental $16^+$ energy, the  moment of inertia ($\mathcal{J}$) should increase as the spin transitions from the low-spin states ($J^\pi$=$0^+$-$6^+$, characterized by a rotational constant $k\approx$78 keV) to the high-spin region. Our SCM calculation
  yields a smaller rotational constant, $k\approx$42 keV, for the high-spin states ($J^\pi$=$12^+$-$16^+$). This value is roughly half of that for the low-spin states, and this decrease in $k$ (which corresponds to an increase in the moment of inertia $\mathcal{J}$), enables the reproduction of the experimental energy of the $16^+$ state.
 }
  This   state has  never been  considered in the context of Bose-Einstein condensation of six $\alpha$ clusters.
Thus, the roton rotational band, $K_{\rm rot}$, of the SCM is considered to  have a dual property of superfluidity  and crystallinity, a property of a supersolid.

\par
In order to confirm that the states calculated in the SCM displayed in Fig.~\ref{fig3} form a well-deformed band structure, the calculated electric transition probabilities are shown in Table 1. We see that the $B(E2)$ value from the $2^+_{n=1}({\rm BdG})$ state to the NG  state $0^+_{\nu=1}({\rm NG})$ is much stronger than that to the vacuum
$0^+_{\nu=0}({\rm NG})$ state. This confirms that the first excited NG  $0^+$  is a band head state of the roton band, $K_{\rm rot}$. The $B(E2)$ values for the states with  $J^\pi\ge$2 of the BdG  excitations show that these states belong to a band with a large deformation consistent with the  $^{12}$C($0_2^+$)+$^{12}$C($0_2^+$)  cluster structure \REV{demonstrated in the geometrical model \cite{Hirabayashi1995}}.
\REVB{It should also be noted that the $B(E2)$ values for the intraband transitions between the roton band states are on the order of 1000 $e^2\text{fm}^4$. This is one order of magnitude greater than the transition between the BdG phonon states, $B(E2: 2^+_{n=1}(\text{BdG}) \rightarrow 0^+_{n=1}(\text{BdG}))$, which is approximately 100 $e^2\text{fm}^4$. This provides further evidence that the roton band states do not represent a vibrational phonon mode, but rather a state characterized by rotational motion.
 }

 \begin{table}[t!]
\caption{ 
$B$({E}2) values (in units of $e^2\,{\rm fm}^4$) and $B$({E}0) values (in units of $e^2\,{\rm fm}^2$) for the ${E}0$ and ${E}2$ transitions in $^{24}$Mg. These values are calculated using the SCM for the  NG and BdG states above the six-$\alpha$ threshold. The calculations are shown for two condensation rates: (a) 70\% and (b) 80\%
}
\label{tab:BE2}
		\begin{tabular}{lclrr} \hline
 		\multicolumn{3}{c}{Transition}  	        		& (a)&  (b)  \\
		 \hline
$B({\rm E}0:0^+_{\nu=1}({\rm NG})$\, &$\rightarrow$ \, &$0^+_{\nu=0}({\rm NG}))$  & 14.1&12.8\\
$B({\rm E}0:0^+_{\nu=2}({\rm NG})$\,&$\rightarrow$ \,& $0^+_{\nu=0}({\rm NG}))$  & 1.3&1.1\\
$B({\rm E}2:2^+_{n=1}({\rm BdG})$\, &$\rightarrow$ \, &$0^+_{n=1}({\rm BdG}))$  & 100&105\\
$B({\rm E}2:2^+_{n=1}({\rm BdG})$\, &$\rightarrow$ \, &$0^+_{\nu=0}({\rm NG}))$  & 1179&1032\\
$B({\rm E}2:2^+_{n=1}({\rm BdG})$\, &$\rightarrow$ \, &$0^+_{\nu=1}({\rm NG}))$  &8365&13609\\
$B({\rm E}2:4^+_{n=1}({\rm BdG})$\, &$\rightarrow$ \, &$2^+_{n=1}({\rm BdG}))$  &
1528 &1567 \\
$B({\rm E}2:6^+_{n=1}({\rm BdG})$\, &$\rightarrow$ \, &$4^+_{n=1}({\rm BdG}))$  & 2168&2192\\
$B({\rm E}2:8^+_{n=1}({\rm BdG})$\, &$\rightarrow$ \, &$6^+_{n=1}({\rm BdG}))$  & 2665&2661\\		
	$B({\rm E}2:10^+_{n=1}({\rm BdG})$\, &$\rightarrow$ \, &$8^+_{n=1}({\rm BdG}))$  & 3044&3020\\		
	$B({\rm E}2:12^+_{n=1}({\rm BdG})$\, &$\rightarrow$ \, &$10^+_{n=1}({\rm BdG}))$  & 3383&3350\\		
	$B({\rm E}2:14^+_{n=1}({\rm BdG})$\, &$\rightarrow$ \, &$12^+_{n=1}({\rm BdG}))$  & 3772&3740\\		
	$B({\rm E}2:16^+_{n=1}({\rm BdG})$\, &$\rightarrow$ \, &$14^+_{n=1}({\rm BdG}))$  & 4290&4260\\		
	$B({\rm E}2:18^+_{n=1}({\rm BdG})$\, &$\rightarrow$ \, &$16^+_{n=1}({\rm BdG}))$  & 4986&4954\\		
	 \hline 
\end{tabular}
\end{table}

\par

\par
Before the observation of BEC of cold atoms \cite{Anderson1995}, in the early 1990s when the 32.5 MeV molecular resonance state was observed, the Hoyle state  $^{12}$C($0_2^+$) was widely believed to have a well-developed three-$\alpha$ linear chain structure \cite{Morinaga1956,Ikeda1968} as well as a weakly coupled $\alpha$+$^8$Be
 structure \cite{Uegaki1977,Uegaki1978,Uegaki1979}, rather than a BEC state.
Taking into account that the Hoyle state is a BEC, the  $^{12}$C($0_2^+$)+$^{12}$C($0_2^+$) cluster structure is considered the first example of a dinuclear BEC molecule
 —a supersolid of two BEC nuclei. 
This may be a prototype of a supersolid with a dinuclear molecule, much like the 
$\alpha$+$\alpha$ structure of $^{8}$Be serves as a prototype in the study of cluster structure, as illustrated by the Ikeda diagram \cite{Ikeda1968,Horiuchi1972}.

 \par
 We discuss why a deformed rotational band, $K_{\rm rot}$,  emerges as the BdG excitation of a superfluid vacuum. Among the BdG excited states with $J^\pi$, states with even parity are lower than those with odd parity because the $\alpha$ clusters are trapped in a harmonic oscillator potential. Among the many even-parity BdG states, those that belong to rotational excitations have the lowest energy, which approaches zero for extremely large systems due to the NG mode caused by the SSB of rotational invariance. Thus, the lowest states with \REV{finite}  angular momentum  ($J^\pi\ne0$)  belong to a rotational band with a deformation. As the deformation becomes larger, the excitation energy becomes smaller. This explains why the BdG states with angular momentum forms a deformed rotational band.
  \REV{On the other hand, from the geometrical picture, it has been systematically shown by Freer and his collaborators \cite{Freer1995A,Freer1997} that even though the deformation costs energy, molecular and cluster structures are favored from the shell model point of view. They demonstrated this by exploring the symmetries of the deformed harmonic oscillator (DHO) for both prolate and oblate deformations, visually demonstrating how the degeneracies of the DHO lead to the spatial nucleon densities expected for $\alpha$-cluster structures \cite{Freer2007,Freer2020,Freer2021}.
  }
 
 \par
 As for the  $J^\pi=0$  states, excitation is caused not only by BdG excitation but also by an NG mode excitation. Among the two, the $0^+$ state excited by the NG mode is lower in energy because it arises from the SSB of global symmetry due to condensation and has zero energy for an extremely large system. Therefore, the lowest BdG excitation $0^+$ state is higher in energy than this first excited  NG mode $0^+$ state. In the present case, as shown in Fig. 1, the first BdG $0^+$  state (with $n=1$) is at $E_{\rm c.m.}$=22.36 MeV, while the NG mode $0^+$  state is located at $E_{\rm c.m.}$=21.05 MeV, which means that the excitation energy of the BdG $0^+$  state from the vacuum is 1.8 times larger than that of the NG mode $0^+$  state. The lowest NG $0^+$  state is deformed due to its orthogonality to the spherical vacuum $0^+$  state.
Thus, the  first excited NG state is also the lowest $0^+$  state with deformation. Namely, the first excited NG $0^+$ state due to the SSB of the global phase is also an NG state with $J^\pi=0$  due to the SSB of rotational invariance. Thus, the NG $0^+$
  state due to the SSB of the global phase and the BdG states with \REV{finite} angular momentum form the lowest rotational band ($K_{\rm rot}$)  just above the vacuum. This mechanism is robust for the excited states of a condensate, which explains why similar bands, known as roton bands, emerge in $^{12}$C, $^{16}$O, and $^{20}$Ne , as  noticed in Ref.~\cite{Ohkubo2025}. 
  
\REVB{The mechanism by which a rotational band emerges on the first NG mode $0^+$ state immediately above the vacuum can be summarized as follows. First, the lowest $0^+$   state is generated as an NG mode due to the SSB of the vacuum's global phase. Second, this NG mode $0^+$ is non-spherical because it must be orthogonal to the spherical vacuum $0^+$ state. This non-sphericity violates rotational invariance, leading to the emergence of a rotational band as a means to restore that invariance.
}   
  This explains also why the roton band is  deformed and  has the  duality of  superfluidity and crystallinity.

Next, we consider why the deformed structure has \REV{dominantly}  the $^{12}\text{C}(0_2^+)+\,^{12}\text{C}(0_2^+)$ configuration. If one $\alpha$ cluster is excited from the six-$\alpha$ vacuum, it forms an $\alpha + \,^{20}\text{Ne}(5\alpha)$ structure. For a two-$\alpha$ cluster excitation, a $^8\text{Be}+\,^{16}\text{O}(4\alpha)$ cluster structure is created. For a three-$\alpha$ cluster excitation, the $^{12}\text{C}(0_2^+)+\,^{12}\text{C}(0_2^+)$ configuration is formed. 
\REV{The six-$\alpha$ breakup threshold in $^{24}\text{Mg}$,  $E_{\text{c.m.}}$=14.55 MeV, is close to the threshold energy for the $\alpha+\,^{20}\text{Ne}(5\alpha$ breakup) channel, which is also 14.55 MeV. In $^{20}\text{Ne}$, a candidate for a five-$\alpha$ condensate state has been observed in Ref. \cite{Adachi2020} at $E_{\text{x}}$=21.2 MeV, 2.0 MeV above the five- $\alpha$ breakup threshold (19.17 MeV).
The threshold energy for both the $^{12}\text{C}(0_2^+)+\,^{12}\text{C}(0_2^+)$ and $^8\text{Be}(0^+)+\,^{16}\text{O}(4\alpha,\,$ 15.10 MeV $ \, 0^+)$  channels is 15.30 MeV. The observed 19.4 MeV state \cite{Fujikawa2024} is 4.1 MeV above the $^{12}\text{C}(0_2^+)+\,^{12}\text{C}(0_2^+)$ threshold and 1.4 MeV below the Coulomb barrier of about 5.5 MeV \cite{Ito2001}.
It is likely that the 19.4 MeV condensate state \cite{Fujikawa2024} observed in the inelastic scattering reaction has \REV{dominantly} a $^{12}\text{C}(0_2^+)+\,^{12}\text{C}(0_2^+)$ configuration rather than the other possible deformed configurations: $^8\text{Be}(0^+)+\,^{16}\text{O}(4\alpha$ 15.10 {MeV} $0^+)$  and $\alpha + \,^{20}\text{Ne}(5\alpha)$, although the mixing of them are likely.
In fact, this structure has a moment of inertia that corresponds to the values obtained
 from the  SCM and cluster model calculations \cite{Hirabayashi1995}.  If the state
  were dominantly composed of the $^8\text{Be}(0^+)+\,^{16}\text{O}$ and  $\alpha +
   \,^{20}\text{Ne}$ configurations, the classically estimated moment of inertia would be
    considerably smaller, in disagreement with the value given in Fig.~\ref{fig3}. The
     moments of inertia for the different configurations change even within the $^{12}$C+$^{12}$C   configuration depending on the excitation energy of $^{12}$C \cite{Abe1980,Hirabayashi1995}.
}

\begin{figure}[t!]
\begin{center}
\includegraphics[width=8.6cm]{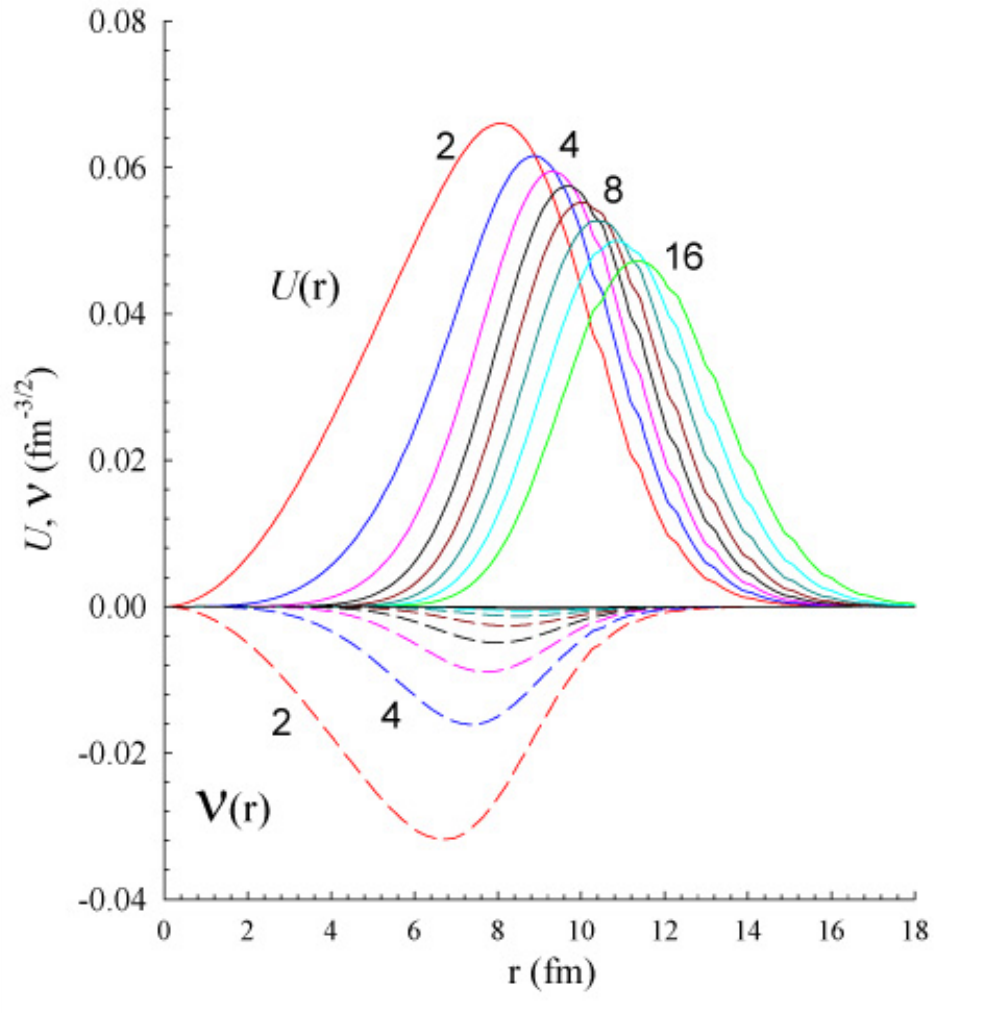}
\caption{
The BdG wave functions, ${\mathcal U}_{n\ell}(r)$ and ${\mathcal V}_{n\ell}(r)$, for the roton band states of the six-$\alpha$ condensate in $^{24}$Mg with  $n=1$ and  $\ell=2$ - 16  under a condensation rate of 70\%
 }
\label{fig:BdGwf}
\end{center}
\end{figure}

\par
\REVB{In the next section, we demonstrate that the roton band states possess a well-developed $^{12}$C($0_2^+$)+$^{12}$C($0_2^+$) cluster structure by calculating the overlap between their wave functions and that of a geometrical $^{12}$C($0_2^+$)+$^{12}$C($0_2^+$) cluster configuration. Before proceeding to the next section, we investigate the properties of the roton-band wave functions. Figure~\ref{fig:BdGwf} displays
}
 the calculated BdG wave functions, $\mathcal{U}_{n\ell}(r)$ and $\mathcal{V}_{n\ell}(r)$ with $n=1$ in Eq.~(\ref{eq:UV}), for the roton band states.
 As  $\ell$ increases, the peak position $R_p$ of ${\mathcal U}_{1\ell}(r)$ moves outward, with  $R_p$ being 8.1, 9.7, and 11.4 fm for $\ell$=2, 8, and 16, respectively. This is reasonable, considering that states with larger $\ell$ correspond to higher excitation energies, as shown in Fig.~\ref{fig3}.
Conversely, Fig.~\ref{fig:BdGwf} shows that  ${\mathcal V}_{1\ell}$(r) is small compared
 to ${\mathcal U}_{1\ell}$(r), especially for large $\ell$. This indicates that the size of
  the condensate is predominantly determined by the behavior of the wave function 
  ${\mathcal U}_{1\ell}$(r). We also observe that the amplitudes of 
  ${\mathcal V}_{1\ell}$(r)  are strongly damped beyond $r=12$ fm in the surface 
    region where ${\mathcal U}_{1\ell}$(r) survives. The magnitude of 
    ${\mathcal V}_{1\ell}$(r), which represents the quantum fluctuations of the
     $\alpha$ clusters of the condensate, decreases as $\ell$ increases. For $\ell$=2,
      the magnitude of  ${\mathcal V}_{12}(r)$ in the internal region is not small compared
       to ${\mathcal U}_{12}(r)$, which means that quantum fluctuations are significant
        for $\ell$=2.
         \REV{We also note that the sizes of the roton band states, ranging from the root-mean-square (rms) radius $R$=8.70 to 12.37 fm, are much larger than that of the vacuum $0^+$ state ($R_0$=6.4 fm), indicating that these states are more dilute in density.
}

\begin{figure}[t!]
\begin{center}
\includegraphics[width=7.6cm]{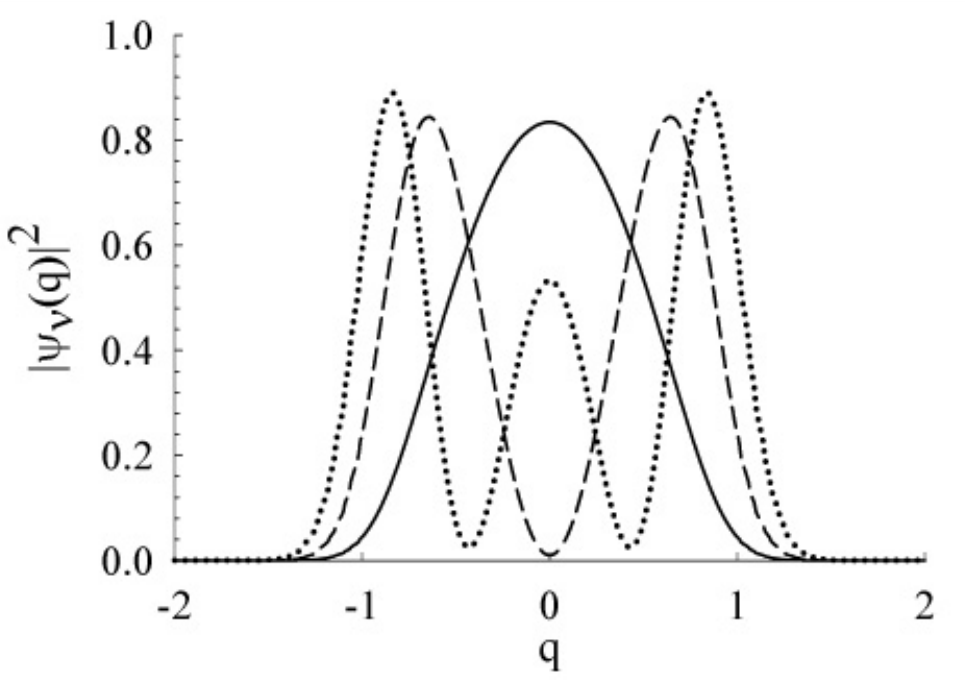}
\caption{
The squares of the wave functions for the NG zero-mode states,  $|\Psi_\nu(q)|^2$
  (for $\nu$=0, 1 and 2), of the six $\alpha$ clusters in $^{24}$Mg, calculated with a condensation rate of 70\%
}
\label{fig:NG-wf}
\end{center}
\end{figure}
 
\begin{figure*}[t!]
\begin{center}
\begin{tabular}{c}
\begin{minipage}{0.330\hsize}
\label{xi}
\begin{center}
\includegraphics[clip, width=5.5cm]{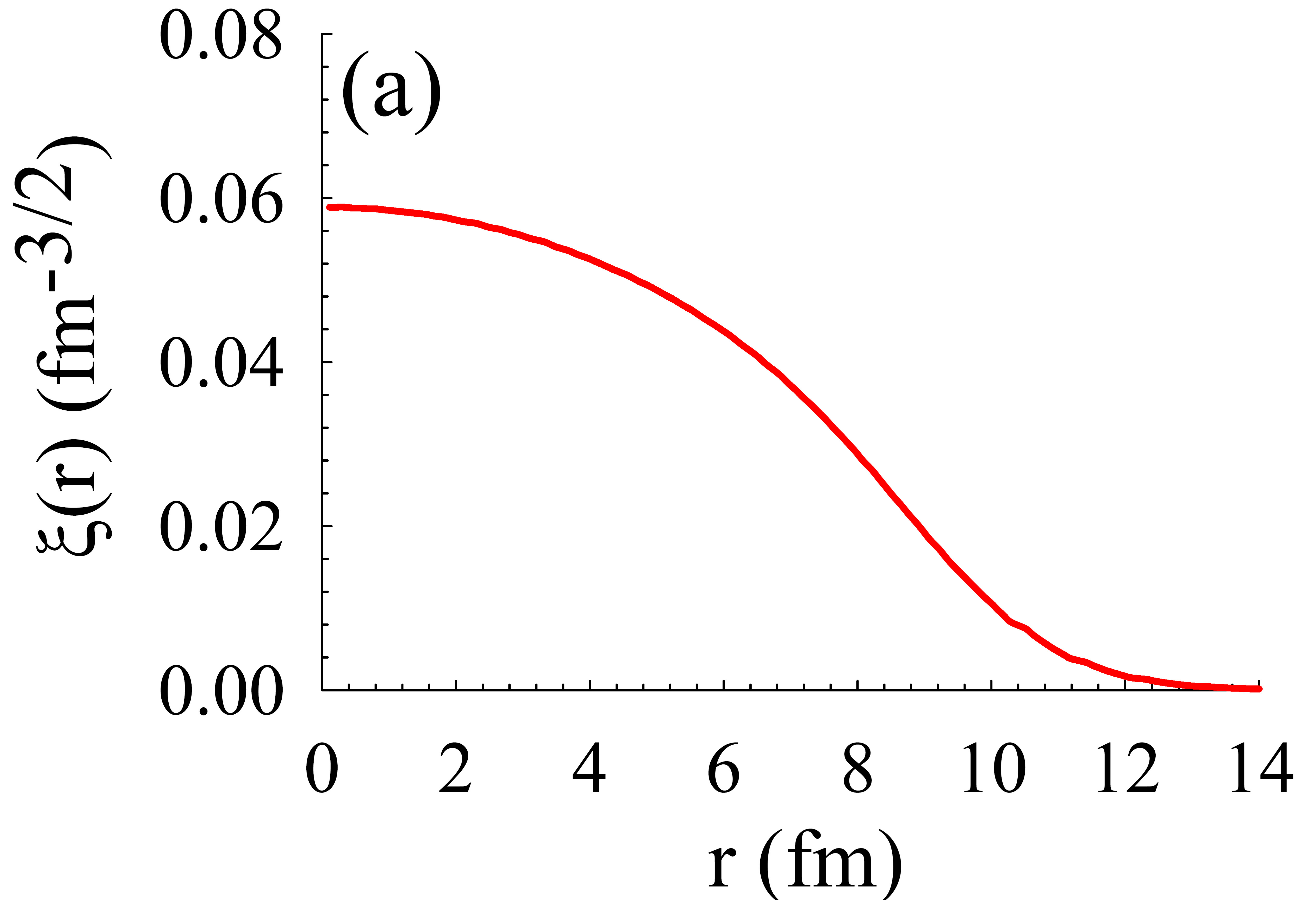}
\hspace{1cm} 
\end{center}
\end{minipage}
\begin{minipage}{0.325\hsize}
\label{eta}
\begin{center}
\includegraphics[clip, width=5.5cm]{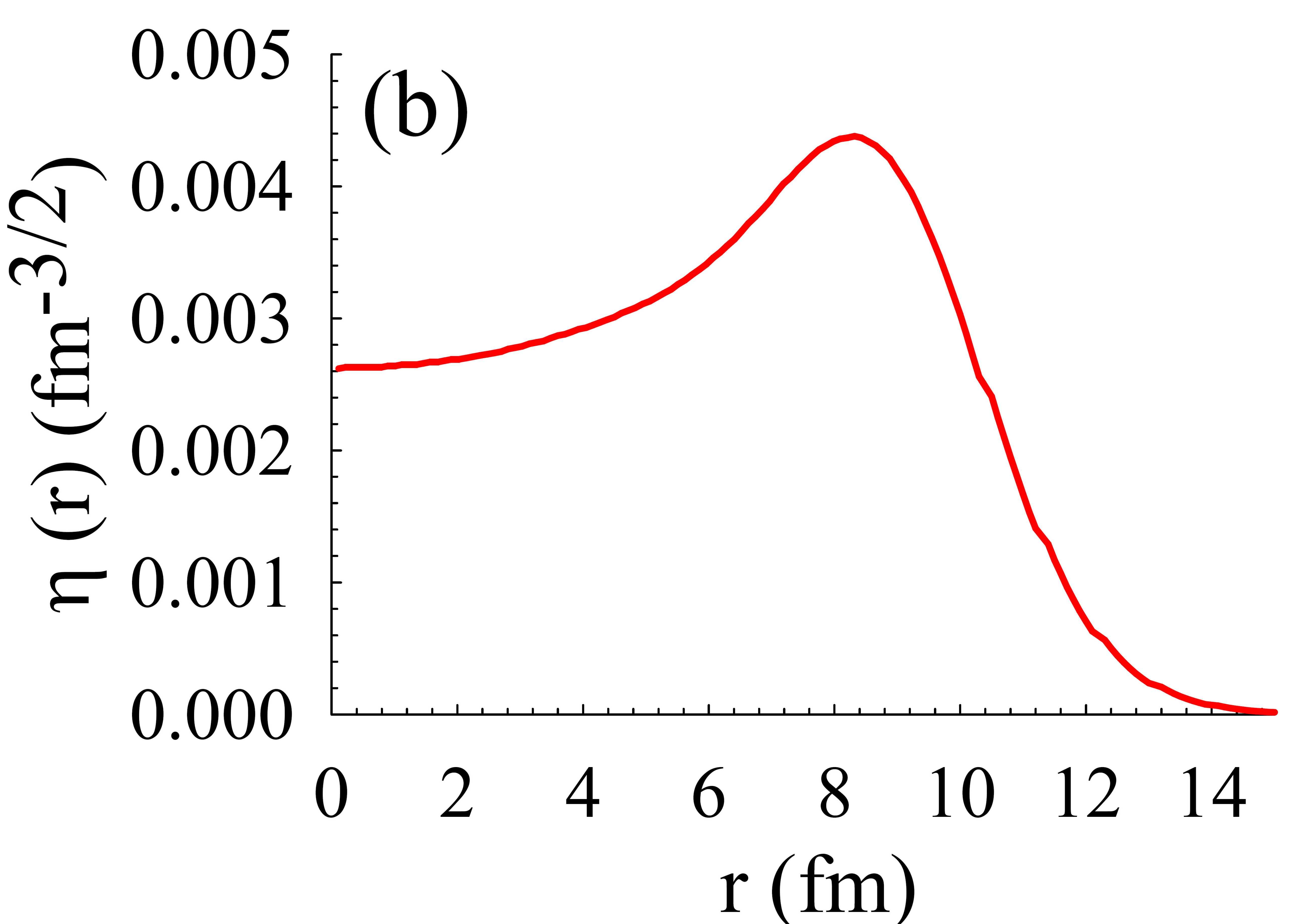}
\hspace{1cm} 
\end{center}
\end{minipage}
\begin{minipage}{0.325\hsize}
\label{density}
\begin{center}
\includegraphics[clip, width=5.5cm]{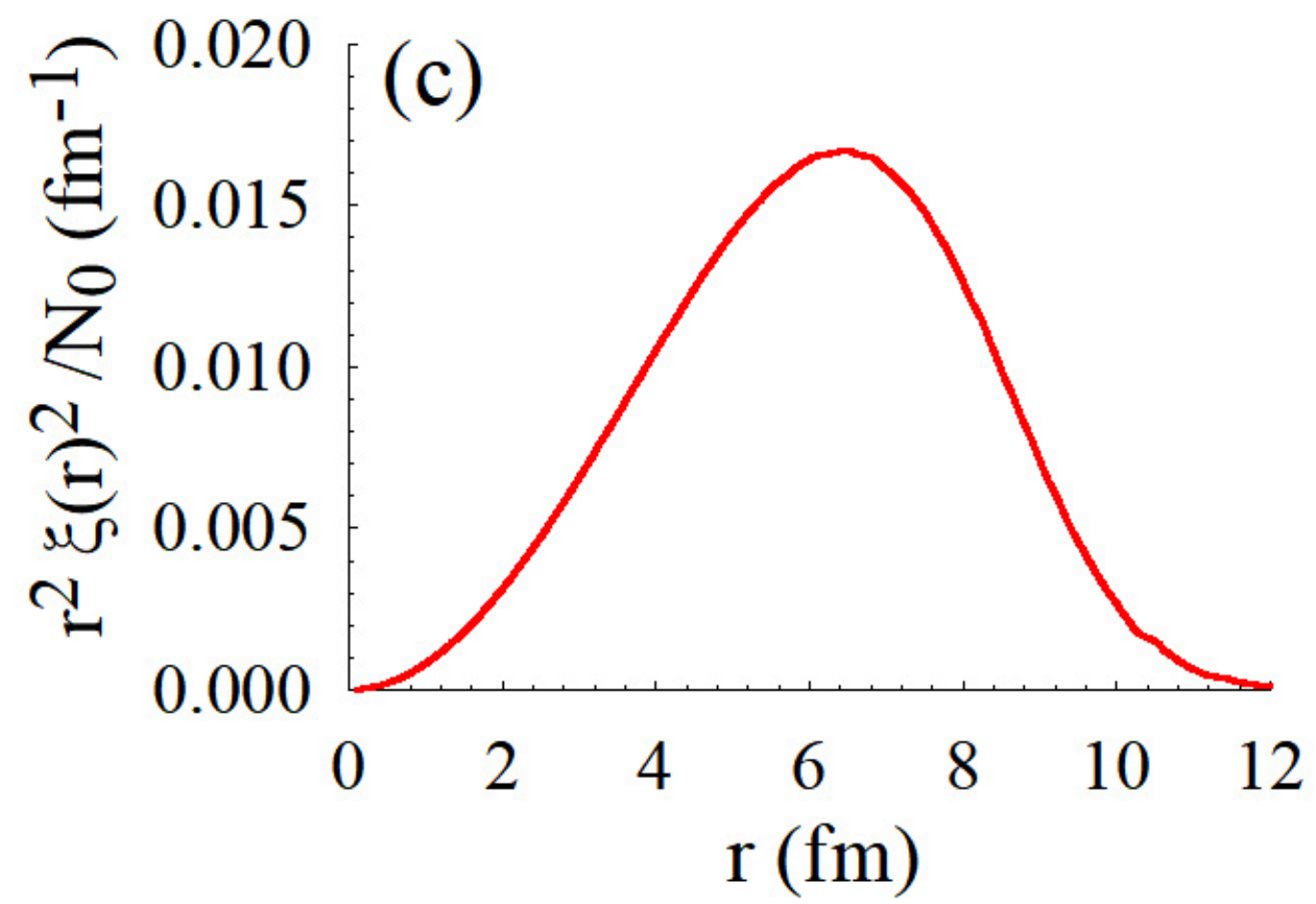}
\hspace{1cm} 
\end{center}
\end{minipage}
\end{tabular}
\end{center}
\caption{ (a) the eigenfunction with zero eigenvalue
$\xi(r)$, order parameter, (b) its adjoint eigenfunction $\eta(r)$, and (c) radial density distribution of the condensate $r^2 | \xi(r)|^2/N_0$ calculated with the condensation rate 70\%  in $^{24}$Mg  }
\label{fig:NG}
\end{figure*}

\par
As for the NG mode $0^+$ states, Fig.~\ref{fig:NG-wf} shows the calculated $|\Psi_{\nu}(q)|^2$
  for $\nu=0\,,\,1\,$, and $\,2$, which are obtained by solving Eq.~(\ref{eq:HuQPeigen}). The nodal excitation of $\Psi_{\nu}(q)$ with respect to $q$ in the NG subspace causes the excitation of the NG mode.
The NG wave functions and their excitation energies depend  little on the size of the condensate, $R_0$ and $\Omega$. This is because the
  functions  $\xi(r)$ and $\eta(r)$ are only present within the integrands of  $\hat H_u^{QP}$, and their integrated values are insensitive to $\Omega$ \cite{Katsuragi2018}. 
 On the other hand,  $\hat H_u^{QP}$ includes  a factor $V_r$, and the NG energy levels rise with an increasing $V_r$\,.  
  The wave function of the vacuum state with  $\nu$=0 has no node. The wave function of the band head $0^+$ state of the roton band ($K_{\rm rot}$), with $\nu$=1, has one node. The state is a collective excitation mode in the gauge space $q$. For an infinite system, the NG zero modes are degenerate with zero energy. However, due to the finite system size and the Pauli principle, the NG mode states form a "band" in the NG subspace $q$. 
 The member states of an NG mode are analogous to the member states of a rotational band of a deformed nucleus. The rotational band is an NG mode that results from the SSB of rotational invariance in a deformed nucleus. In this analogy, an NG mode member state  resulting from the SSB of global phase with quantum number  $\nu$ and eigenenergy $E_\nu$ corresponds to a rotational state with angular momentum $I$ and eigenenergy $E(I)$. 
 
\par
In Fig.~\ref{fig:NG} (a), the eigenfunction  $\xi(r)$ with a zero eigenvalue, which serves as the order parameter, is shown. 
The quantity $|\xi(r)|^2$ represents the condensate fraction density.
In Fig.~\ref{fig:NG} (b), the adjoint eigenfunction $\eta(r)$, calculated as a derivative of $\xi(r)$ with respect to the number of $\alpha$ clusters, $N_0$, represents the fluctuation of the number of $\alpha$ clusters.
 In Fig.~\ref{fig:NG} (c), the radial density distribution of the condensate, defined by $ r^2 \xi^2(r)/N_0$, is displayed. We see from Fig.~\ref{fig:NG} (c) that the superfluid density distribution of the condensate $\alpha$ clusters in the vacuum state extends to about 12 fm, which is consistent with the picture of a diffused, gas-like structure.
In Fig.~\ref{fig:NG} (b), we note that the number fluctuation around the average $N_0$
  slowly increases from the center at  $r=$0  fm with a depression, reaching its largest
   peak at $r=8$ fm, and then decreases toward the surface, extending to about 14 fm.
    This indicates that the number fluctuation occurs not only in the central region but
     also in the surface region, beyond the rms radius of the
      condensate, which is 6.4 fm.
      
  \par
 We would like to mention the robustness of the roton band's emergence \REV{on the condensation rates.} Our calculations so far have been performed by assuming a 70\% condensation rate. The calculated band for an 80\% condensation rate is indistinguishable from the one presented in Fig.~\ref{fig3}. Moreover, the transition rates for 80\% condensation, as listed in Table 1, also show strong $B(E2)$ values between the states of the roton band. \REV{The dependence of the energy levels and the $B(E2)$ values on the condensation rate is weak. } Therefore, the emergence of the roton band with the six-$\alpha$ condensation, exhibiting a duality of superfluidity and crystallinity in $^{24}$Mg, is robust. This is similar to the
  condensations of three-$\alpha$, four-$\alpha$, and five-$\alpha$ in $^{12}$C, $^{16}$O, and $^{20}$Ne, respectively.

  \section{     Supersolidity with  \REVB{ $^{12}$C($0_2^+$)+$^{12}$C($0_2^+$)} dinuclear Bose-Einstein condensate
   }\label{sec:5}
 
 \REVB {
To examine the $^{12}$C($0_2^+$)+$^{12}$C($0_2^+$) clustering features of the SCM roton band, we calculate the overlap of the SCM wave functions  ${\mathcal U}_{1\ell}$(r)  with the geometrical cluster wave functions $\psi_\ell(r, d)$. The latter is a normalized wave function defined by two $^{12}$C($0_2^+$) clusters separated by an rms distance $d$. We adopt the condensate wave function for $^{12}$C($0_2^+$) from Refs.~\cite{Nakamura2016,Katsuragi2018}, which has an rms radius of 3.8 fm and a 70\% condensation rate, and has successfully described the BEC-like $\alpha$ cluster states in $^{12}$C, such as the band built on the first  NG $0^+$ state \cite{Ohkubo2025}.
   }

 \par
\begin{figure}[t!]
\begin{center}
\includegraphics[width=7.8cm]{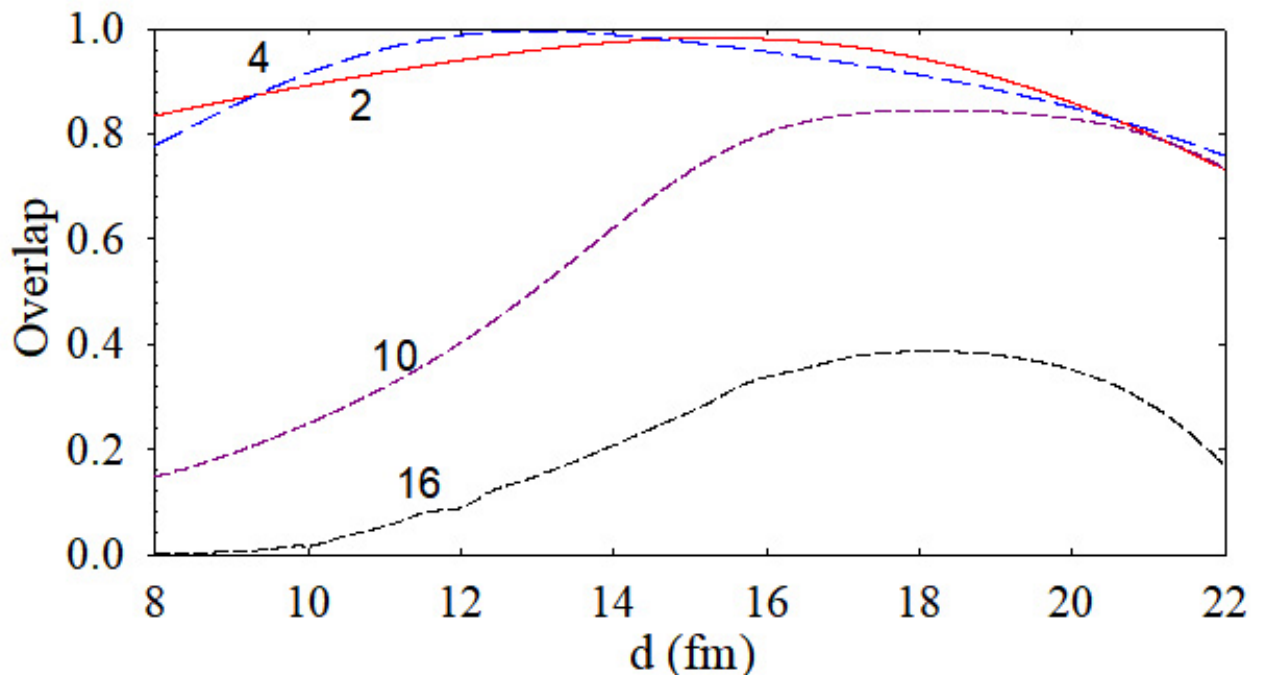}
\caption{ \REVB{  
Overlap integrals calculated between the SCM roton band wave functions and the geometrical $^{12}$C($0_2^+$)+$^{12}$C($0_2^+$) cluster wave functions, plotted as a function of the intercluster distance $d$ between the c.m. of the two $^{12}$C($0_2^+$)  clusters. The labels indicate the spin $J$ of the roton band states.
}
}
\label{fig:overlap}
\end{center}
\end{figure}

    \REVB{     
  Fig.~\ref{fig:overlap} shows the calculated overlap as a function of $d$. It is evident that the overlaps for the $2^+$ and $4^+$ states are near-unity in the range $d$ = 10–20 fm (e.g., 0.98 for $2^+$ and 0.97 for $4^+$ at $d \approx 15$ fm). This provides strong evidence that the $2^+$ and $4^+$ states of the roton band possess a geometrical $^{12}$C($0_2^+$)+$^{12}$C($0_2^+$) cluster structure. While the overlap values generally decrease as the spin of the roton band increases, they remain as high as 0.84 for the $10^+$ state (at $d \approx 9$ fm) and 0.39 for the $16^+$ state (at $d \approx 18$ fm). This indicates that the SCM wave functions retain a significant $^{12}$C($0_2^+$)+$^{12}$C($0_2^+$) cluster component even at high spin. As seen in Fig.~\ref{fig:overlap}, the degree of clustering—measured by the optimal distance $d$—generally increases with higher spin. Although this trend appears to be inverted for the $2^+$ and $4^+$ states (where the maximum overlap is 0.983 at $d = 15.4$ fm for $2^+$ and 0.996 at $d = 13.0$ fm for $4^+$), it should be noted that the overlap for $4^+$ remains as high as 0.96 at $d = 16.0$ fm. Indeed, near-unity overlap values are maintained throughout the $10 \le d \le 20$ fm region. As discussed in the following paragraphs, this persistence of high overlap is a clear signature of the supersolidity of the roton band states, characterized by a geometrical cluster structure coexisting with superfluidity.
    }

  \REVB{
   We can also demonstrate another distinctive characteristic of the SCM wave functions: a cluster structure characterized by an extraordinarily large intercluster distance. In conventional cluster model calculations of the $^{12}$C($0_2^+$)+$^{12}$C($0_2^+$) system—which has been studied extensively over the past few decades in the context of molecular resonances at high excitation energies in $^{24}$Mg \cite{Hirabayashi1995,Ito1999,Ito2001}—double-folding model analyses show that the $^{12}$C($0_2^+$)+$^{12}$C($0_2^+$) potential vanishes beyond $r \approx 10$ fm (see Figs. 4 and 5 of Ref. \cite{Ito1999}). In contrast, our results indicate that the $^{12}$C($0_2^+$)+$^{12}$C($0_2^+$) structure extends to $d = 10$–$20$ fm, a range inaccessible to ordinary geometrical cluster models. Indeed, the system forms a configuration with an exceptionally large separation where the clusters can move freely, consistent with the picture of superfluidity. Consequently, the $2^+$ and $4^+$ states of the roton band possess extremely large and highly dilute components. This extended nature is also evident in higher spin states, such as the $10^+$ and $16^+$ states of the roton band, where the maximum overlap occurs in the even more extended region of $d = 17$–$20$ fm.
  }  
       
   \REVB{
  The characteristic that SCM wave functions exhibit a large and significant overlap over a wide range of intercluster distances is a clear indication of superfluidity. In a conventional geometrical configuration with a definite distance $d_0$, the overlap would peak at $d_0$ and decrease sharply at larger and smaller distances. However, it should be noted that the overlap remains significantly large throughout the entire $d = 10$–$20$ fm region. This indicates that while the SCM wave functions possess a $^{12}$C($0_2^+$)+$^{12}$C($0_2^+$) cluster structure consistent with the conventional model, they simultaneously incorporate cluster wave functions spanning a broad distribution of distances $d$. The fact that the overlap is nearly constant over such a wide range implies that the intercluster distance can vary with negligible energy cost. Consequently, the two $^{12}$C($0_2^+$) clusters can move almost freely within this range—a hallmark of superfluidity analogous to superfluid $^4$He (He II), where atoms flow without viscosity (i.e., without energy dissipation). This situation persists for the higher spin states of the roton band, although the extent of the plateau region in the overlap decreases with increasing spin. Thus, it is evident that the SCM wave functions simultaneously manifest both a geometrical $^{12}$C($0_2^+$)+$^{12}$C($0_2^+$) structure and superfluidity, which is the defining property of supersolidity.
      }

  \REVB{
 To verify the large intercluster distances ($d$) associated with the dinuclear structure of the six-$\alpha$ BEC states in the roton band, we calculate the distance between the two clusters as follows:
}
  The intercluster distance $d$ between the c.m.
 of the two clusters, C1 and C2, is calculated using the following equation:
\REVR{ \begin{equation}
M \langle R^2 \rangle = \frac{ m_1 m_2}{M} \langle d^2 \rangle + m_1 \langle R^2_1 \rangle  +  m_2 \langle R^2_2 \rangle .
\label{eq:rms}
\end{equation}
}
Here, \REVR{$R_1$, $R_2$,} and $R$ denote the rms radii for cluster C1 (with mass \REVR{$m_1$}), cluster C2 (with mass \REVR{$m_2$}), and the total system (with \REVR{$M=m_1+m_2$)}, respectively.
The rms radius $R$ for 
\REVB{the roton band states}
 with angular momentum $\ell$ is calculated using the eigen wave functions of the BdG excitations,  $\mathcal{U}_{n\ell}(r)$ and $\mathcal{V}_{n\ell}(r)$, with $n=1$, as shown in Fig.~\ref{fig:BdGwf}.
  The  rms radius
  of the Hoyle state has not been measured experimentally. Theoretical calculations report slightly different   values depending on the models: 2.89 fm \cite{Danilov2009}, 3.3 fm \cite{Kanada2007}, 3.38 fm \cite{Chernykh2007}, 3.47 fm \cite{Fukushima1977,Kamimura1981}, 3.83 fm \cite{Tohsaki2001,Schuck2004,Funaki2008}, 4.28 fm \cite{Matsumura2004}, and 4.31 fm \cite{Yamada2004,Yamada2005}. All these values are larger than the rms radius of the ground state of   $^{12}$C (2.47 fm) due to the dilute nature of the Hoyle state.
 
  In Table \ref{tab:rms}, the intercluster distance $d$ between the two dinuclear  $^{12}$C($0_2^+$)  clusters is calculated for four selected model cases: (a) 2.89 fm \cite{Danilov2009}, (b) 3.38 fm \cite{Chernykh2007}, (c) 3.83 fm \cite{Tohsaki2001}, and (d) 4.31 fm \cite{Yamada2004,Yamada2005}. The intercluster distance increases with the spin $J$ of the roton rotational band, indicating the enhancement of separation with increasing $J$. For $J$=2, $d$ is in the range of 15-16 fm, and for $J$=16, $d$ is in the range of 22-23 fm. 
\REVB{These $d$ values for the $^{12}$C($0_2^+$)+$^{12}$C($0_2^+$) cluster structure are quantitatively consistent with the results presented in the preceding paragraphs. Furthermore, it is noteworthy that these values are similar to those obtained from other model calculations, such as the 18 fm length of the six-$\alpha$ linear chain estimated in the eigenshape model and the 24 fm length estimated in the six-$\alpha$ cluster model \cite{Rae1992}. This similarity is particularly interesting given that the underlying physical pictures of these models are fundamentally different.
}

\REV{We have also calculated the intercluster distance $d$ for the $^8\text{Be}(0^+)+\,^{16}\text{O}(4\alpha,\,$15.10 MeV $ \, 0^+)$ structure and  the $\alpha$ + $^{20}$Ne(5$\alpha$) structure. For the first case, we assumed a size of 2.53 fm for the ground state of $^8$Be, which has a long lifetime with a width of 5.57 eV. This size is obtained from Eq.~(\ref{eq:rms}) using the  $\alpha$ particle size of 1.676 fm and the distance of 3.5 fm between the two  $\alpha$ particles in the cluster model \cite{Horiuchi1970}. The size of the  $^{16}\text{O}(4\alpha,\,$ 15.10 MeV $ \, 0^+)$ state is assumed to be 5.60 fm. The obtained values for the intercluster distance $d$ range from 15.4 fm for $J=2$ to 24.2 fm for $J=18$. For the second case, the $\alpha + \,^{20}\text{Ne}(5\alpha)$ structure, we assumed a size of 6.0 fm for $^{20}\text{Ne}(5\alpha)$ and obtained $d$ values ranging from 18.0 fm for $J=2$ to 29.7 fm for $J=18$. Thus, regardless of the cluster structure considered, our states in the roton band with a deformed structure possess a well-separated cluster configuration. As shown in the preceding paragraphs,  }
\REVB{the  $^{12}$C($0_2^+$) + $^{12}$C($0_2^+$)  cluster structure is dominant, although mixing with the $^8\text{Be}(0^+)+\,^{16}\text{O}(4\alpha,\,$ 15.10 MeV $ \, 0^+)$  structure remains possible. }

 \begin{table}[t!]
\caption{ 
The calculated rms radius $R$ of the six-$\alpha$ roton band states with spin $J(=\ell)$ and the intercluster rms distance $d$ (in units of fm) between the two Hoyle $^{12}$C$(0_2^+)$ clusters. The distance $d$ is calculated for the reported rms radius 
 of the Hoyle $^{12}$C$(0_2^+)$ state from different theoretical models: (a) $2.89\ \text{fm}$ \cite{Danilov2009}, (b) $3.38\ \text{fm}$ \cite{Chernykh2007}, (c) $3.83\ \text{fm}$ \cite{Tohsaki2001}, and (d) $4.31\ \text{fm}$ \cite{Yamada2004,Yamada2005}
}
\label{tab:rms}
\begin{tabular}{cccccc}
		 \hline
	$J$ &  $R$ & \multicolumn{4}{c}{ intercluster distance $d$} \\
           & & (a) & (b)  & (c)  & (d)  \\	
 		  \hline
		 $2^+$ & 8.70    & 16.41 &16.03 &  15.62 &   15.11 \\
		 $4^+$ & 9.31    & 17.70 &17.35 &  16.97 &   16.50 \\		
		 $6^+$ & 9.74    & 18.60 &18.27 &  17.91 &   17.47 \\
		  $8^+$ & 10.10  & 19.36 &19.04 &  18.69 &   18.27  \\
		 $10^+$ &10.46   & 20.11 &19.80 &  19.47 &   19.06 \\		
		 $12^+$ & 10.85  & 20.92 &20.62 &  20.30 &   19.91 \\	 
		  $14^+$ & 11.30 & 21.85 &21.57 &  21.26 &   20.89 \\
		 $16^+$ & 11.81  & 22.90 &22.63 &  22.34 &   21.99 \\		
		 $18^+$ & 12.37  & 24.06 &23.80 &  23.52 &   23.19 \\	 
	 \hline 
\end{tabular}
\end{table}

 \par
\REVB{ Having demonstrated that the roton band states exhibit a well-developed $^{12}$C($0_2^+$)+$^{12}$C($0_2^+$) cluster structure—representing a supersolid phase of dinuclear BEC condensates—we finally discuss the possibility of the Josephson effect.  While this effect has been observed in various macroscopic systems, including cold atom systems \cite{Gati2007,Albiez2005,Ryu2013,Javanainen1986}, its manifestation in finite nuclear systems has primarily focused on Cooper pair transfer between superfluid heavy nuclei, such as $^{116}$Sn and $^{60}$Ni \cite{Broglia2022,Potel2021}. 
}
In contrast, the Josephson effect between $\alpha$-condensates has been largely overlooked. The $^{12}$C($0_2^+$)+$^{12}$C($0_2^+$) configuration may provide a \REVB{ unique} opportunity to investigate a Josephson current driven by $\alpha$ condensation. Although the $^{12}$C($0_2^+$) state is unstable above the $\alpha$-decay threshold, its decay width ($\Gamma_{\rm c.m.} = 9.3$ eV) is extremely narrow. This long lifetime is sufficient to permit the transfer of a\REVB{ condensed boson} between the two nuclei. Let $\phi_1$ and $\phi_2$ be the phases of the macroscopic wave functions of the two $^{12}$C($0_2^+$) condensate nuclei. The global phase difference, $\Delta\phi = \phi_1 - \phi_2$, could \REVB{ induce a Josephson current between the two $^{12}$C($0_2^+$) clusters. Experimental evidence supports this picture.} A resonance in the $^{16}$O+$^8$Be system has been observed at an almost identical energy, showing a similar angular distribution but with a narrower width \cite{Aliotta1995,Freeman1995}. At the resonance energy and 90$^\circ$, the cross section for the inelastic $^{12}$C($0_2^+$)+$^{12}$C($0_2^+$) channel is 35 $\mu$b/sr, whereas the $^{16}$O+$^8$Be(g.s.) channel reaches 120–170 $\mu$b/sr \cite{Aliotta1995,Freeman1995}. The decay of the 32.5 MeV resonance into the $^{16}$O+$^8$Be channel with angular momentum $\ell = 16$ was further confirmed in Ref. \cite{Szilner1997}. \REVB{Notably, } the excitation spectrum of $^{12}$C($^{12}$C,$^8$Be)$^{16}$O (Fig. 1(b) in Ref. \cite{Freeman1995}) shows that the cross section to the excited 15.1 MeV $0^+$ state of $^{16}$O is the strongest—more than five times larger than other decay channels. This enhancement is consistent with the fact that the 15.1 MeV $0^+$ state is considered a four-$\alpha$ condensate \cite{Funaki2008C,Takahashi2020}, while the excited state $^{16}$O($0_2^+$, 6.05 MeV) has a well-developed $\alpha$ + $^{12}$C(g.s.) cluster structure and the $^{16}$O($0_3^+$, 11.26 MeV)  state has a  [$\alpha(L=2)$$\otimes$ $^{12}$C($2^+$)]$_{J=0}$  cluster structure \cite{Suzuki1976A,Suzuki1976B}  with a large $\alpha$ decay width of 2.5 MeV \cite{Tilley1993}. Here, $L$ is the orbital angular momentum between $\alpha$ and $^{12}$C.

Given that the threshold energy for the $^{12}$C($0_2^+$)+$^{12}$C($0_2^+$) channel (15.30 MeV) is nearly degenerate with that of the $^{16}$O(15.1 MeV)+$^8$Be channel (15.30 MeV), these data strongly suggest that the $^{12}$C($0_2^+$)+$^{12}$C($0_2^+$) resonance structure coexists with the $^8$Be+$^{16}$O(15.1 MeV) structure at 32.5 MeV. The structural transition from $^{12}$C($0_2^+$)+$^{12}$C($0_2^+$) to $^8$Be+$^{16}$O(15.1 MeV) can be qualitatively understood as a one-$\alpha$ transfer process. This transfer is triggered by the phase difference $\Delta\phi$ of the Josephson effect. \REVB{As shown in Sect. 5, the intercluster distance $d$ ranges from 16 to 22 fm for $J^\pi = 2^+$ to $16^+$ and the two condensate  $^{12}$C($0_2^+$) clusters move with negligible  energy cost.} Although these separations are large, the wave function of the $^{12}$C($0_2^+$) state extends up to $r \approx 8$ fm \cite{Uegaki1979}, allowing the tails of the two wave functions to overlap. Consequently, a Josephson current of $\alpha$ clusters between the two Hoyle-like nuclei in the roton band becomes a plausible mechanism. \REVB{The observation of  $\gamma$ rays due to charged $\alpha$ transfer could provide evidence for such a process. While this discussion remains speculative, it underscores the need for more quantitative theoretical and experimental exploration of the cluster structure in roton band states.}
 
\section{Summary}\label{sec:6}
\par
We have \REVB{demonstrated,} for the first time, that the recently observed candidate states for a six-$\alpha$ condensate with low spins ($J \le 4^+$) in $^{24}$Mg \REVB{are} well described by the superfluid $\alpha$-cluster model. This model rigorously treats the Nambu-Goldstone (NG) zero mode as the order parameter of condensation for the finite six-$\alpha$ system. It is found that a roton rotational band with a large moment of inertia is built on  the first excited NG $0^+$ state, \REVB{analogous} to the roton bands observed in three-, four-, and five-$\alpha$ condensates in $^{12}$C, $^{16}$O, and $^{20}$Ne, respectively. The calculated roton band successfully reproduces the \REVB{well-known} molecular resonance with a $^{12}$C($0_2^+$)+$^{12}$C($0_2^+$) structure at $E_{\rm c.m.} = 32.5$ MeV, characterized by a spin of $16^+$, as observed in inelastic $^{12}$C+$^{12}$C scattering. 
\REVB{Consequently,} the low-spin six-$\alpha$ condensate states located slightly above the vacuum and the $^{12}$C($0_2^+$)+$^{12}$C($0_2^+$) molecular resonances are now understood within \REVB{a single, unified framework}. 
\REVB{Furthermore, the roton band is found to exhibit a duality of superfluidity and crystallinity, being dominantly composed of a two-cluster Hoyle-like $^{12}$C($0_2^+$)+$^{12}$C($0_2^+$) structure. This coexistence of properties is a definitive signature of a supersolid.
These results suggest that the duality of crystallinity and superfluidity may be a common feature across multi-$\alpha$ systems, encouraging further  studies on supersolidity in both light and heavy nuclei.
The present discovery  suggests a new link between nuclear cluster physics and the universal study of quantum liquids, such as ultracold atomic systems, opening future prospects for exploring supersolidity across different energy scales.
}

\begin{itemize}
\item Funding
No funding.
\item Conflict of interest/Competing interests 
Not applicable
\item Consent for publication
The authors agree with publication of the manuscript in the traditional publishing model.
\item Data availability Statement
This manuscript has no associated data or the data will not be deposited. 
[Authors’ comment: This is a theoretical study using  published data, and all data information is properly referenced.]
\item Code availability Statement
This manuscript has no associated code/software. 
[Authors’ comment: Code/Software sharing not applicable to this article as no code/software was generated or analyzed during  the current study.]

\end{itemize}

\end{document}